\documentclass[letterpaper, 10 pt, conference]{ieeeconf}  

\IEEEoverridecommandlockouts                              

\usepackage{booktabs} 
\usepackage{url}
\usepackage{subfig}
\usepackage{graphicx}
\graphicspath{{figures/}}
\usepackage[english]{babel}
\usepackage[autostyle=true,english=american]{csquotes}
\usepackage{gensymb}
\usepackage{supertabular}

\def\sectionautorefname{Section}

\usepackage{acro}
\DeclareAcronym{UI}{short = UI, long = user interface}
\DeclareAcronym{GUI}{short = GUI, long = graphical user interface}
\DeclareAcronym{TLX}{short = NASA TLX, long = NASA-Task Load Index}
\DeclareAcronym{RTLX}{short = NASA Raw-TLX, long = NASA Raw-Task Load Index}
\DeclareAcronym{ER}{short = ER, long = error rate}
\DeclareAcronym{TCT}{short = TCT, long = task completion time}
\DeclareAcronym{HCI}{short = HCI, long = human-computer interaction}
\DeclareAcronym{UX}{short = UX, long = user experience}
\DeclareAcronym{RMSE}{short = RMSE, long = root mean squared error}
\DeclareAcronym{HMD}{short = HMD, long = Head-Mounted Display}
\DeclareAcronym{CNN}{short = CNN, long = Convolutional Neural Network}
\DeclareAcronym{FOV}{short = FoV, long = field of view}
\DeclareAcronym{HRC}{short = HRC, long = Human-Robot Collaboration}
\DeclareAcronym{HRI}{short = HRI, long = Human-Robot Interaction}
\DeclareAcronym{ANOVA}{short = ANOVA, long = analysis of variance}
\DeclareAcronym{RMANOVA}{short = RM-ANOVA, long = Repeated Measures Analysis of Variance}
\DeclareAcronym{JND}{short = JND, long =just-noticeable difference}
\DeclareAcronym{SUS}{short = SUS, long =system usability scale}
\DeclareAcronym{CSCW}{short = CSCW, long = computer-supported cooperative work}
\DeclareAcronym{CAD}{short = CAD, long = computer-aided design}
\DeclareAcronym{MR}{short = MR, long = Mixed Reality}
\DeclareAcronym{AR}{short = AR, long = Augmented Reality}
\DeclareAcronym{AV}{short = AV, long = Augmented Virtuality}
\DeclareAcronym{VR}{short = VR, long = Virtual Reality}
\DeclareAcronym{SAR}{short = SAR, long = Spatial Augmented Reality}
\DeclareAcronym{ADLs}{short = ADLs, long = Activities of Daily Living}
\DeclareAcronym{LED}{short = LED, long = Light-Emitting Diode}
\DeclareAcronym{DoF}{short = DoF, long = Degree-of-Freedom, long-plural-form = Degrees-of-Freedom}
\DeclareAcronym{HHC}{short = HHC, long = Human-Human Collaboration}
\DeclareAcronym{AI}{short = AI, long = Artifical Intelligence}
\DeclareAcronym{QUEAD}{short = QUEAD, long = Questionnaire for the Evaluation of Physical Assistive Devices}
\DeclareAcronym{TiA}{short = TiA, long = Trust in Automation Questionnaire}
\DeclareAcronym{ADMC}{short = ADMC, long = Adaptive DoF Mapping Control, long-plural-form = Adaptive DoF Mapping Controls}

\title{\LARGE \bf
In Time and Space: Towards Usable Adaptive Control for Assistive Robotic Arms
}

\author{Max Pascher$^{1,2}$ and Kirill Kronhardt$^{1}$ and Felix Ferdinand Goldau$^{3}$ and Udo Frese$^{3}$ and Jens Gerken$^{1}$
\thanks{$^{1}$Max Pascher, Kirill Kronhardt, and Jens Gerken are with the Westphalian University of Applied Sciences, Human-Computer Interaction, 45897 Gelsenkirchen, Germany
{\tt\small max.pascher@w-hs.de, kirill.kronhardt@w-hs.de, jens.gerken@w-hs.de}}%
\thanks{$^{2}$Max Pascher is also with the University of Duisburg-Essen, Human-Computer Interaction, 45127 Essen, Germany
{\tt\small max.pascher@uni-due.de}}%
\thanks{$^{3}$Felix Ferdinand Goldau and Udo Frese are with the University of Bremen, Mathematics \& Computer Science, 28359 Bremen, Germany
{\tt\small fgoldau@uni-bremen.de, ufrese@uni-bremen.de}}%
}

\begin{document}

\maketitle
\thispagestyle{empty}
\pagestyle{empty}


\begin{abstract}
Robotic solutions, in particular robotic arms,  are becoming more frequently deployed for close collaboration with humans, for example in manufacturing or domestic care environments.
These robotic arms require the user to control several \acp{DoF} to perform tasks, primarily involving grasping and manipulating objects. Standard input devices predominantly have two DoFs, requiring time-consuming and cognitively demanding mode switches to select individual DoFs. Contemporary \acp{ADMC} have shown to decrease the necessary number of mode switches but were up to now not able to significantly reduce the perceived workload. Users still bear the mental workload of incorporating abstract mode switching into their workflow. We address this by providing feed-forward multimodal feedback using updated recommendations of ADMC, allowing users to visually compare the current and the suggested mapping in real-time. 
We contrast the effectiveness of two new approaches that a) \emph{continuously} recommend updated DoF combinations or b) use discrete \emph{thresholds} between current robot movements and new recommendations. Both are compared in a \ac{VR} in-person study against a \emph{classic} control method. Significant results for lowered task completion time, fewer mode switches, and reduced perceived workload conclusively establish that in combination with feedforward, ADMC methods can indeed outperform classic mode switching. A lack of apparent quantitative differences between \emph{Continuous} and \emph{Threshold} reveals the importance of user-centered customization options. 
Including these implications in the development process will improve usability, which is essential for successfully implementing robotic technologies with high user acceptance. 
\end{abstract}



\section{Introduction}
\label{sec:introduction}


\noindent While robotic devices have long been put behind fences for safety reasons, advances in the development of such (semi-) autonomous technologies have started to permeate almost all aspects of our personal and professional lives. These include increased close-quarter collaborations with robotic devices -- from industry assembly lines~\cite{Braganca.2019} to mobility aides~\cite{Fattal.2019}. Assistive robotic arms are a particularly useful and versatile subset of collaborative technologies with varied applications in different fields, e.g.,~\cite{Pulikottil.2021,Beaudoin.2018}.

Yet, new challenges arise when robots are tasked with (semi-) autonomous actions, resulting in additional stress for end-users if not correctly addressed during the design process~\cite{Pollak.2020}. Pollak et al. highlight the decreased feeling of control users experienced when using a robot's autonomous mode. Switching to manual mode allowed their study participants to regain control and decrease stress significantly. These findings are corroborated by Kim et al. whose comparative study of control methods resulted in markedly higher user satisfaction for the manual mode cohort~\cite{Kim.2012}.


A proposed solution from previous work~\cite{Goldau.2021petra} to these challenge are adaptive controls -- referred as Adaptive DoF Mapping Controls (ADMCs) -- which merge the advantages of (semi-) autonomous actions with the flexibility of manual controls. They combine multiple \acp{DoF} dynamically for a specific scenario to assist in controlling the robot. 
In our concept, a \ac{CNN} interprets a camera's video feed of the environment and dynamically combines the most likely \acp{DoF} for a suggested movement. Building on this, we already showed that such \ac{ADMC} combinations of the robot's \acp{DoF} can lead to a significantly lower number of mode switches compared to standard control methods~\cite{Kronhardt.2022adaptOrPerish}. However, our study could not show that this may also improve task completion time or reduce cognitive load. Also, challenges concerning the understanding of \ac{DoF} mappings were raised during the study. 


Based on these previous findings, the present study evaluates two novel \acp{ADMC} methods for an assistive robotic arm. We compare the variants \emph{Continuous} and \emph{Threshold}, differing in the time at which suggestions are communicated to the user, against a \emph{classic} control method. In detail, we examine possible effects on task completion time, number of necessary mode switches, perceived workload, and subjective user experience.
Our contribution is two-fold:
\begin{enumerate}
    \item We demonstrate that both \ac{ADMC} methods significantly reduce the task completion time, the average number of mode switches, and the perceived workload of the user.
    \item Further, we establish that for \emph{Continuous} and \emph{Threshold}, each has specific advantages which some users may prefer over the other, raising the need for customizable configurations.  
\end{enumerate}
\section{Related Work}
\label{sec:related}

\noindent Collaborative robotic solutions have received much attention in recent years. Previous work has generally focused on (a) 
different designs of robot motion intent and most recently (b) \acp{ADMC} for robots. The latter requires a critical yet seldom addressed topic in how collaborative robots can effectively communicate recommended movement directions to their user.

\subsection{Robot Motion Intent}
\label{sec:background-robotIntents}
\noindent Advance knowledge of the intended robot behavior and subsequent movements within the physical world are critical for effective collaboration when humans and robots occupy the same space and need to coordinate their actions~\cite{walker.2018.armotionintent}. In previous work, we analyzed existing techniques of communicating robot motion intent and identified different \emph{intent types} as well as several intent properties, such as \emph{location} and \emph{information} or the placement of the technology~\cite{Pascher.2023robotMotionIntent}. Users generally prefer to have the robot's future movements represented visually~\cite{cleaver.2021}. To convey detailed robot motion intent, researchers often rely on \ac{AR}~\cite{Hetherington_2021,Chadalavada.2020,Pascher.2022.Perception}, as \enquote{with the help of \ac{AR}, interaction can become more intuitive and natural to humans}~\cite{makhataeva2020}.


Effective communication of robot motion intent is particularly relevant when using \acp{ADMC} for assistive robotic arms, as in such a shared or traded control environment each interaction needs to be precisely coordinated. 





\subsection{Adaptive \ac{DoF} Mapping Controls} 
\label{sec:background-adaptiveControl}
\noindent Traditionally, robot control methods include individual commands for each \ac{DoF}, requiring frequent mode switches for controlling translations, rotations, and gripper functionality. 
Herlant et al. called into question the suitability of these standard control methods as task completion time markedly increases by using user-initiated compared to time-optimal mode switches~\cite{Herlant.2016modeswitch}. 


To tackle this issue, we proposed in previous work the concept of \ac{ADMC} -- a dynamic combination of multiple \acp{DoF}, thus adjusted to specific scenarios or tasks~\cite{Goldau.2021petra}. This streamlining decreases the need for constant mode switching, resulting in faster and more efficient task fulfillment. In~\cite{Goldau.2021petra} we implemented a \ac{CNN} as control unit to provide these dynamic \ac{DoF} mappings and gave the user a triggering mechanism to request an update. In a 2D simulation study which had a 4-\ac{DoF} robot control mapped to a 2-\ac{DoF} input device, we found promising results.



We then extended this approach into a 3D \ac{VR} simulation, thereby mapping a 7-\ac{DoF} robot control to a 2-\ac{DoF} input device~\cite{Kronhardt.2022adaptOrPerish}. We evaluated two \ac{ADMC} methods -- differing in their respective movement suggestion concept -- against the baseline control method \emph{Classic}. Simulating the effect of a \ac{CNN}, our work relied on a task-specific script to provide \ac{DoF} mappings based on the relative position and orientation between gripper and target. This removed the potentially confounding effect of a suboptimal \ac{CNN} implementation. Results showed that the number of mode switches was significantly reduced compared to \emph{Classic}, but task completion time was unaffected. Users reported high cognitive demand and difficulties understanding the mapping to 2 different input \acp{DoF}. In addition, the system felt difficult to predict and required trial and error~\cite{Kronhardt.2022adaptOrPerish}.



\section{Adaptive \ac{DoF} Mapping Controls}
\label{sec:control-types}
\noindent Building on our previous work~\cite{Kronhardt.2022adaptOrPerish}, we created a \ac{VR} simulation of a \ac{HRI} experimental setup to compare different \ac{ADMC} methods to a non-adaptive baseline condition \emph{Classic}. Like in previous work~\cite{Kronhardt.2022adaptOrPerish} we applied a task-specific script to explore our \ac{ADMC} methods. We tackle previous issues by 1) visualizing not only the current but also the forthcoming \ac{DoF} mapping suggestion (improving predictability) and 2) reducing the input to a single \ac{DoF} (reducing cognitive demand). We propose two approaches as different trade-offs between control fidelity and cognitive demand.


The \ac{VR} simulation includes a virtual model of the \emph{Kinova Jaco 2}\footnote{Kinova Robotic arm. \url{https://assistive.kinovarobotics.com/product/jaco-robotic-arm}, last retrieved \today.} -- a commercially available assistive robotic arm frequently used in \ac{HRI} studies, e.g.,~\cite{Beaudoin.2018,Herlant.2016modeswitch}.
Our proposed visual feedback mimics \ac{AR}, with directional cues registered in 3D space. This allows the user to understand different movement directions for the actual control and the suggested \ac{DoF} combinations. To simplify understanding, we use \emph{arrows}, a straightforward and common visualization technique to communicate motion intent~\cite{walker.2018.armotionintent, shrestha.2016.motionintent,shindev.2012.intentexpression}.

As a control method for the \acp{ADMC}, we implemented a task-specific script. This removed any potential bias that a more generic but currently still technically limited approach such as a \ac{CNN}-based control method may introduce. Of course, our approach only works in a controlled experimental setting. The task-specific script evaluates the gripper's current position, rotation, and finger position relative to a target. The \ac{DoF} mapping system then suggests five different movement options (referred in the following to as \emph{modes}) -- in order of assumed usefulness -- to the user.

\begin{enumerate}
    \item \emph{Optimal Suggestion:} Combining translation, rotation, and finger movement [opening and closing] into one suggestion, causing the gripper to move towards the target, pick it up, or release it on the intended surface.
    \item An orthogonal suggestion based on (1) but excluding the finger movement. Allows the users to adjust the gripper's position while still being correctly orientated.
    \item A pure translation towards the next target, disregarding any rotation.
    \item A pure rotation towards the next target without moving the gripper.
    \item Opening or closing of the gripper's fingers.
\end{enumerate}

\begin{figure}[htbp]
\centering
\captionsetup{justification=centering}
    \subfloat[]{\includegraphics[width=0.3195\linewidth]{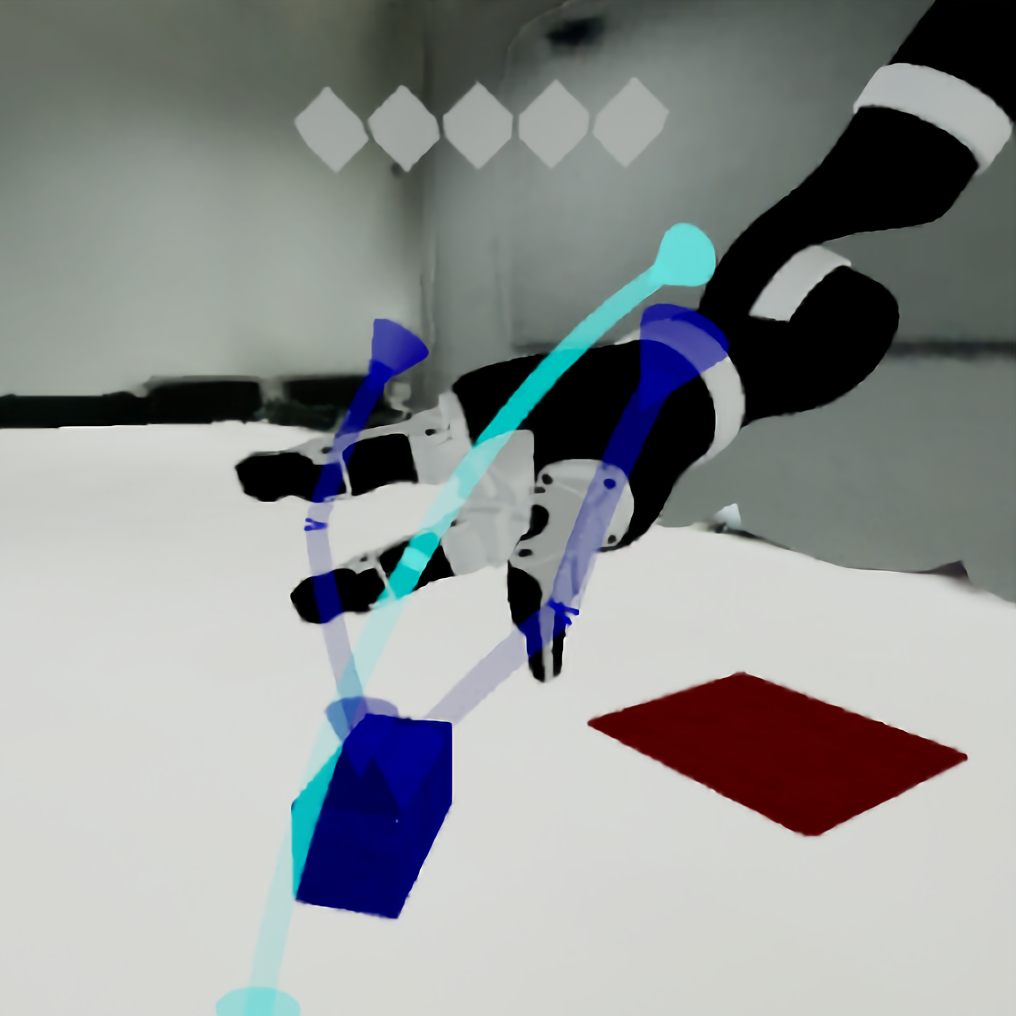}\label{fig:adaptive-suggestions:a}}
    \hfill
    \subfloat[]{\includegraphics[width=0.3195\linewidth]{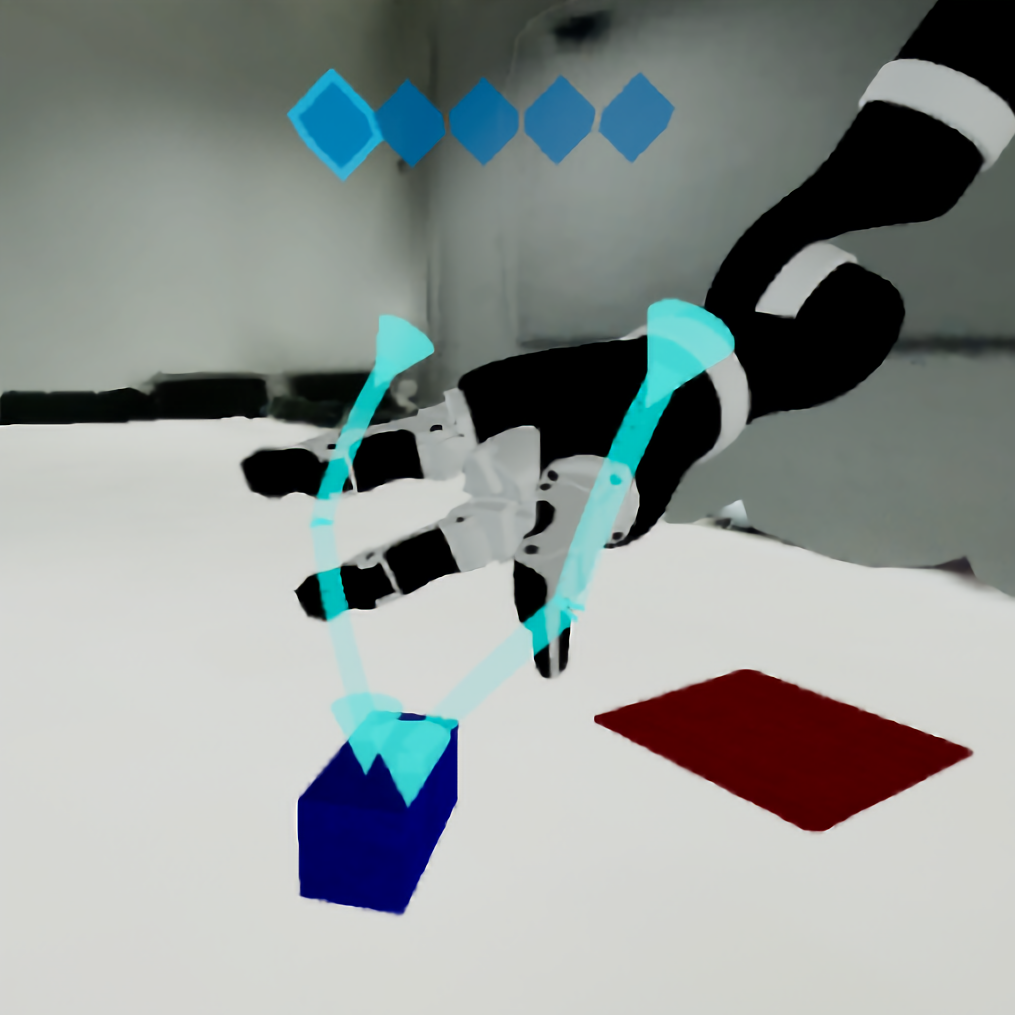}\label{fig:adaptive-suggestions:b}}
    \hfill
    \subfloat[]{\includegraphics[width=0.3195\linewidth]{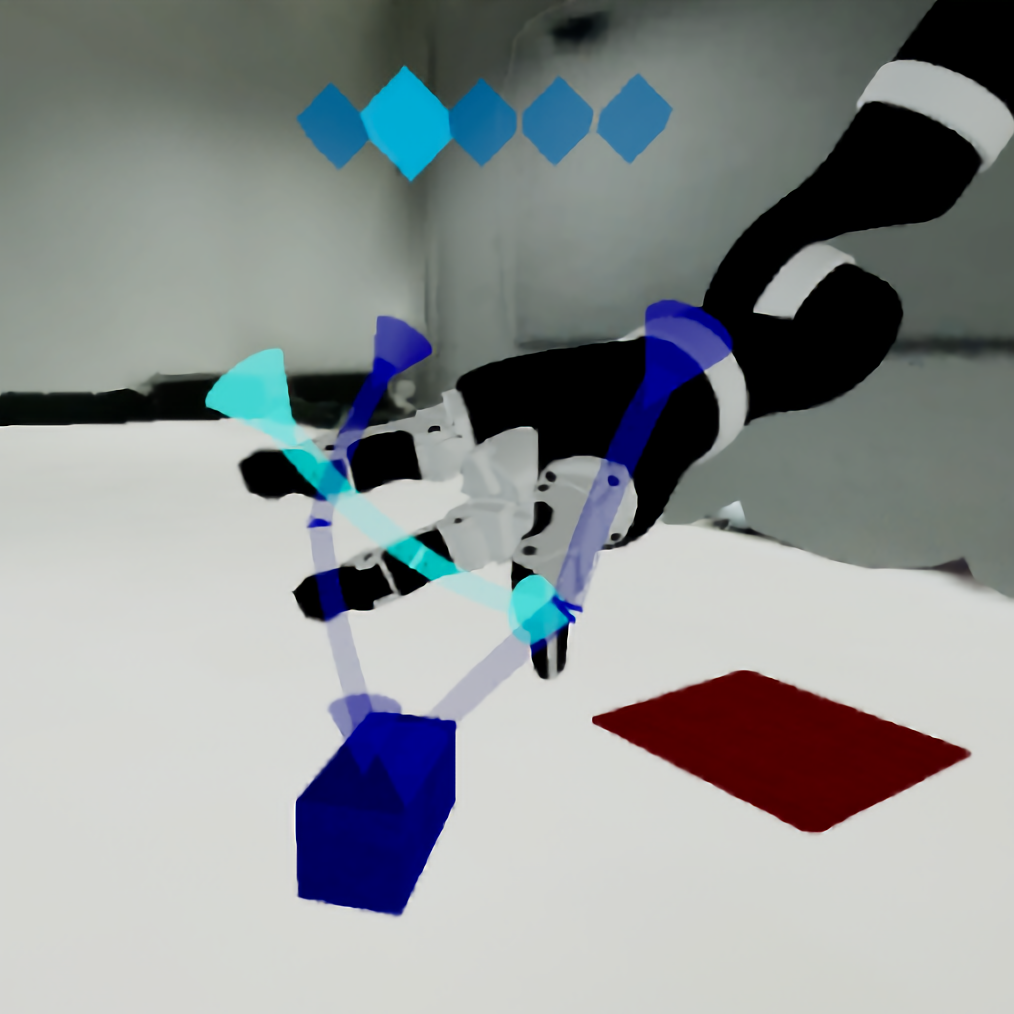}\label{fig:adaptive-suggestions:c}}
    \hfill
    \subfloat[]{\includegraphics[width=0.3195\linewidth]{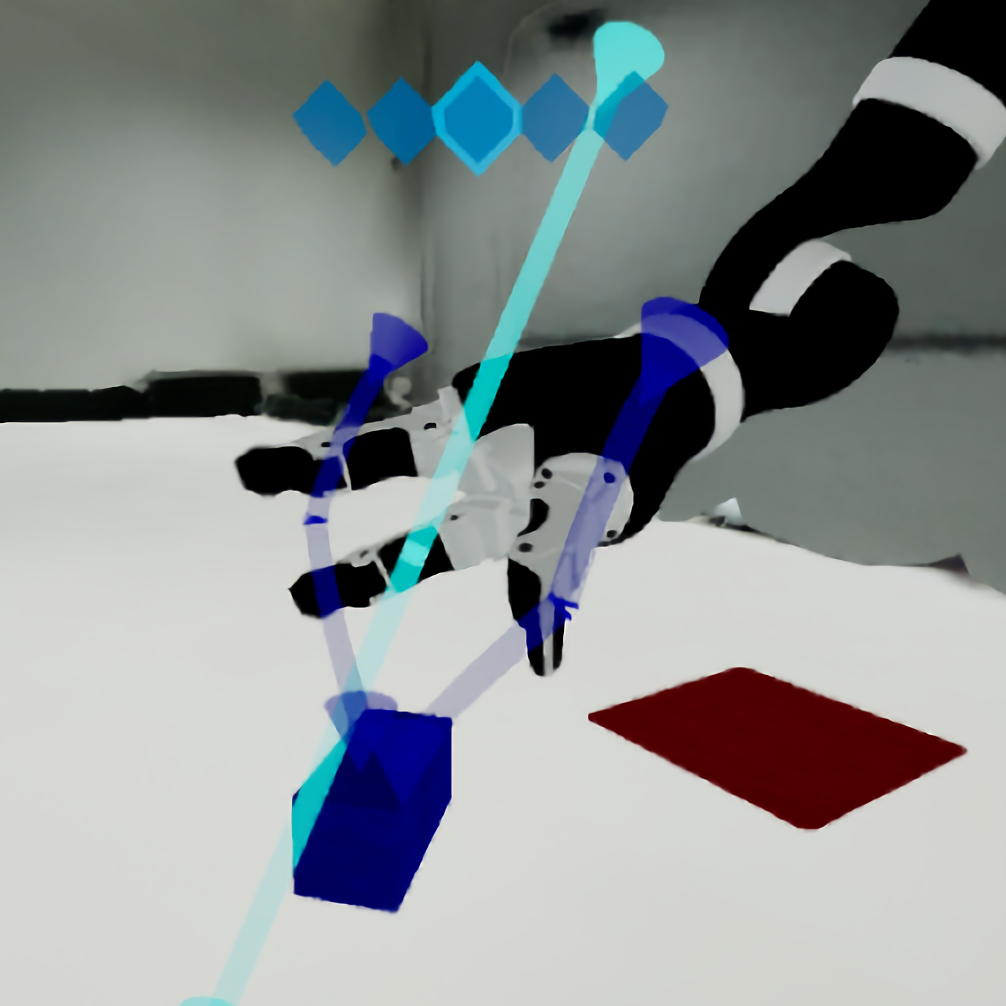}\label{fig:adaptive-suggestions:d}}
    \hfill
    \subfloat[]{\includegraphics[width=0.3195\linewidth]{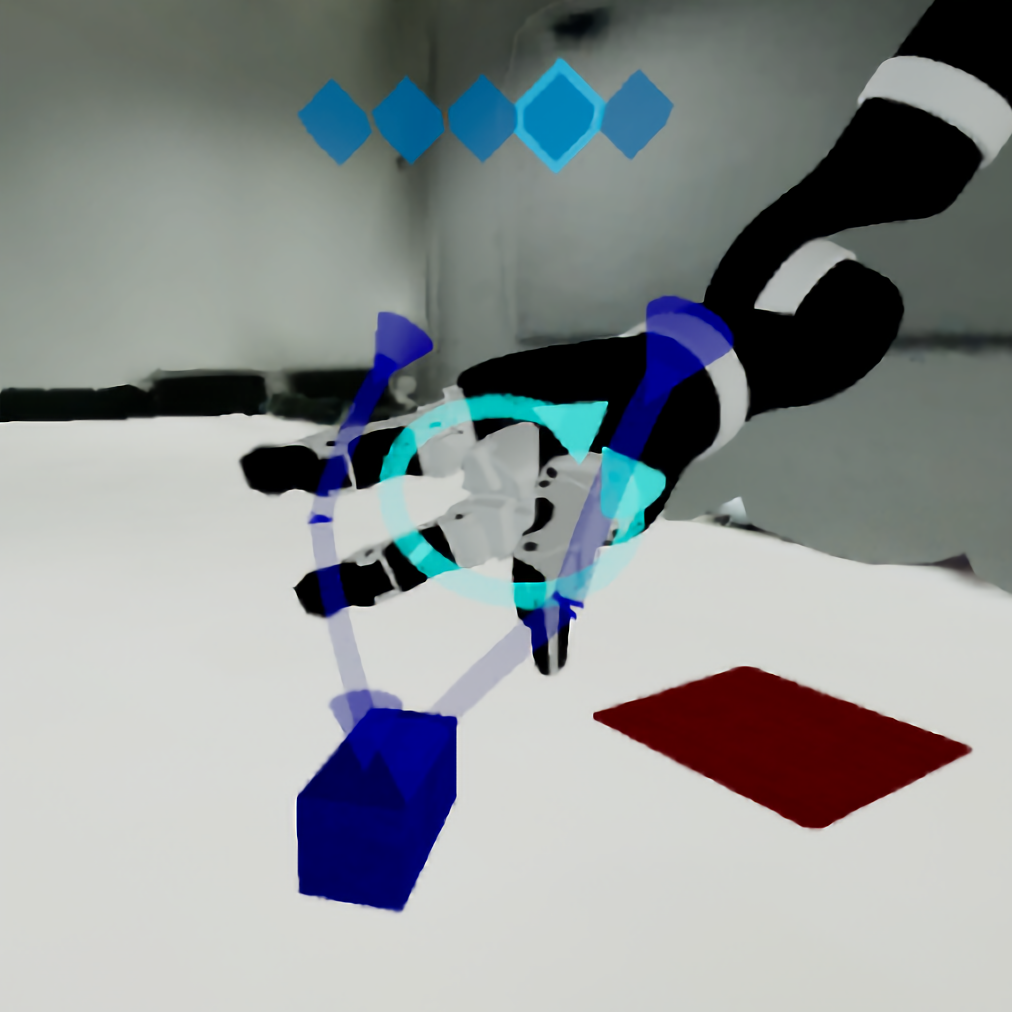}\label{fig:adaptive-suggestions:e}}
    \hfill
    \subfloat[]{\includegraphics[width=0.3195\linewidth]{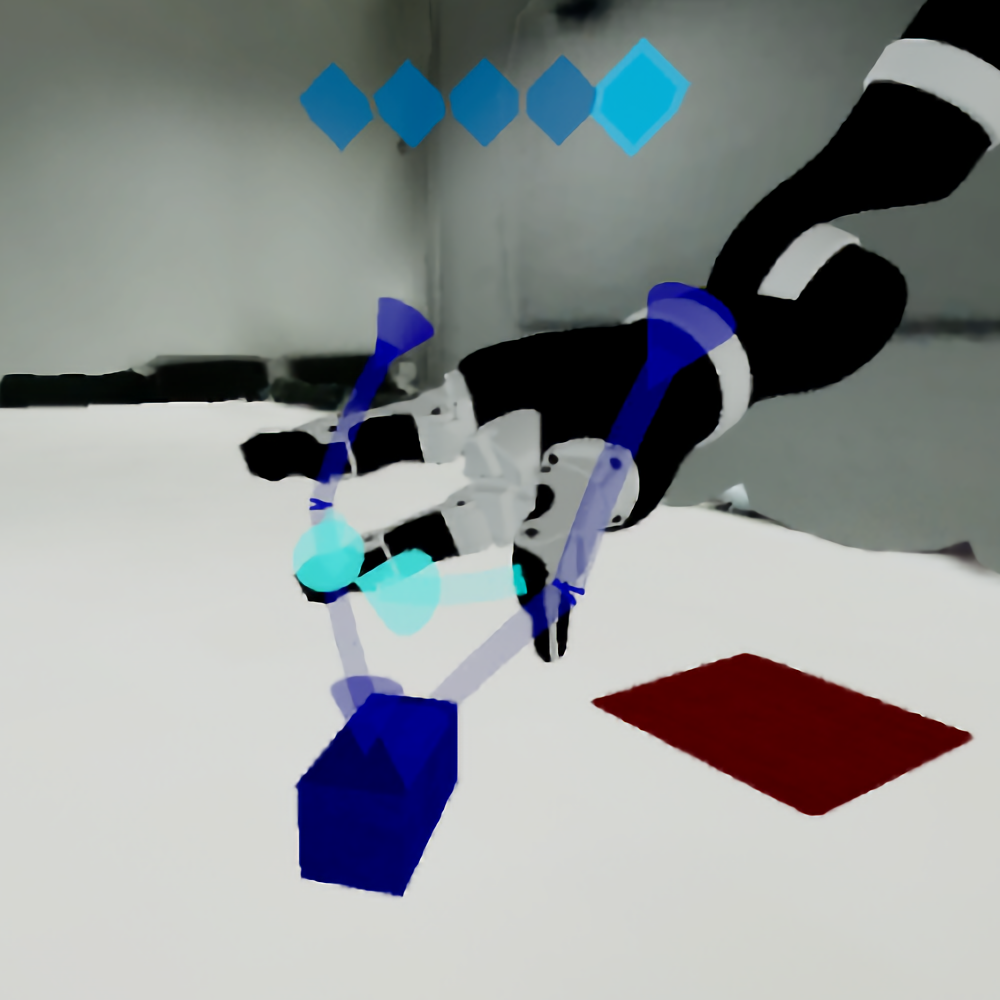}\label{fig:adaptive-suggestions:f}}
\captionsetup{justification=justified}
  \caption{Suggestions as visualized in the \ac{ADMC} methods, (\textbf{a}) Continue previous movement, (\textbf{b}) Optimal Suggestion, (\textbf{c}) Adjustment Suggestion, (\textbf{d}) Pure Translation, (\textbf{e}) Pure Rotation, (\textbf{f}) Open / Close Fingers.}
  \label{fig:adaptive-suggestions}
\end{figure}

\noindent During movement, the \ac{ADMC} system re-calculates the best \ac{DoF} combinations to fulfill the specific task, which are then presented as new suggestions. Users cycle through these modes -- by pressing a button on the input device -- to select a suitable one or continue moving with the previous active suggestion (see Figure~\ref{fig:adaptive-suggestions}). A suggestion indicator is visible above the gripper when users are not moving the robot to distinguish between the modes. Five slanted cubes represent the possible suggestions. The cubes appear gray if no suggestion is active and turn blue to indicate that a new suggestion is selected. The cube corresponding to the selected mode increases in size. 
In contrast to our previous work~\cite{Kronhardt.2022adaptOrPerish} and to the dual axis system of the baseline control method (see Figure~\ref{fig:apparatus}), only one input axis is required to control the robotic arm. Consequently, the cognitive demand on the users is reduced as they can focus on evaluating one movement rather than two simultaneous suggestions.



\noindent\textbf{Continuous}: This control method uses continuous feedback of robot motion intent to increase oversight of updated movement suggestions. Continuous feedback enables users to move in a direction and constantly evaluate the updated optimal suggestion by the \ac{ADMC} system. If found fitting, users can switch to a new suggestion and move the robot in the updated path to fulfill the task. Here, two directional indicators are virtually attached to the robotic arm's gripper: a light blue and a dark blue arrow. The former represents the currently selected movement option (mode) mapped to the input axis. The forward movement of the input axis moves the gripper in the direction the arrow is pointing; engaging it backward moves the gripper in the arrow's reverse direction. The dark blue arrow represents the currently optimal suggestion at a given time. Users can only move the robot along the dark blue arrow if they switch to that suggestion first -- which causes both arrows to overlap.
While this approach increases transparency, users might be distracted by the constantly updating suggestions, potentially leading to more mode switches and perceived workload.

\noindent\textbf{Threshold}: In contrast to \emph{Continuous}, \emph{Threshold} uses time-discrete and multimodal feedback to indicate optimized movement suggestions. Again, a light blue arrow maps the selected movement option (mode) to the input axis. New suggestions are only shown to the users if the optimal mode differs -- by a set degree -- from the current movement. We followed Singhal et al. and used a cosine between-vector similarity measure to calculate this threshold~\cite{singhal2001modern}, ranging from exact alignment [0\%] to total opposite direction [100\%].
In pretests, we determined a 20\% difference between the current and optimal vector as the suggestion threshold. If exceeded, a short vibration pulse to the input device and a 1kHz sound inform the users of an updated suggestion. In addition, a dark blue arrow appears which visualizes the new suggested movement. Users can continue the active movement, switch to the new suggestion, or cycle through the other four modes before deciding on one. Unlike with \emph{Continuous}, users can therefore entirely focus on the movement they are currently performing until explicitly notified and directed to a new suggestion. 
We expect \emph{Threshold} to reduce perceived workload compared to \emph{Continuous} as it does not require constant evaluation of the visual feedback. However, we expect task completion time to increase, as \emph{Threshold} systematically interrupts the users' workflow. Additionally, \emph{Threshold} might result in a perceived loss of control, potentially negatively influencing usability.
\section{Study Method and Materials}
\noindent To explore the effectiveness of our \ac{ADMC} methods, we conducted a supervised, controlled experiment as a \ac{VR} simulation study with 24 participants. We compared our \ac{ADMC} methods to \emph{Classic}, which relies on mode switching to access and control all \acp{DoF} one after another. Approaches as \emph{Classic} are well established (e.g., when driving a car) and are predictable and transparent for the user. Comparing \ac{ADMC} methods to \emph{Classic} allows \ac{HRI} researchers to disentangle their respective advantages and disadvantages. 

\subsection{Study Design}
\label{sec:study-design}
\noindent We applied a within-participant design with \emph{control method} as an independent variable with three conditions: (1) \emph{Classic}, (2) \emph{Continuous}, and (3) \emph{Threshold}. Every participant performed eight training trials and 24 measured trials per condition, resulting in 72 measured and 24 training trials per participant and 1,728 measured trials in total. To counter learning and fatigue effects, the order of conditions was fully counter-balanced.
We measured the following dependent variables: 

\begin{enumerate}
\item \textbf{Average Task Completion Time} For each trial, we measured the time in seconds needed to pick an object and place it on the target surface. 

\item \textbf{Average Number of Mode Switches} For each trial, we recorded every mode switch conducted by pressing a button on the input device.

\item \textbf{Perceived Workload} After completing each condition, we measured cognitive workload with the \ac{RTLX} questionnaire~\cite{Hart.2006}.

\item \textbf{Subjective Assessment} After completing each condition, we measured the five dimensions of the \ac{QUEAD}~\cite{schmidtler2017questionnaire}. After completing all trials, participants were further asked to rank the three conditions.
\end{enumerate}

\noindent After each condition, participants were prompted with several open questions regarding their experience, their understanding of the control methods and the directional cues, plus any issue of interest they considered noteworthy. Additionally, participants were asked how they proceeded in situations when they could not solve the task at first. 

\noindent Video and audio recordings of the interviews with the entire study cohort were assessed independently by two researchers. Open coding was applied to gather participants' opinions of the different control methods. We used Miro\footnote{Miro. \url{https://miro.com}, last retrieved \today.} -- an online whiteboard~\cite{harboe2015realworld} -- to complete an affinity diagram of the open codes. Codes were then organized into themes (see \sectionautorefname~\ref{sec:qualitativeInsights}). 

\subsection{Hypotheses}
\noindent Overall, we expected \ac{ADMC} methods to reduce not just mode switches (as in prior work~\cite{Kronhardt.2022adaptOrPerish}) but -- due to the advances in our designs -- also improve on task completion time and workload. 

\begin{enumerate}
 \item[\textbf{H1:}]  \emph{Continuous} and \emph{Threshold} lead to a lower task completion time compared to \emph{Classic}. However, we expect \emph{Continuous} to perform faster compared to \emph{Threshold}, as the latter systematically interrupts the user during interaction.
 \item[\textbf{H2:}] \emph{Continuous} and \emph{Threshold} result in fewer mode switches compared to \emph{Classic}. We expect \emph{Continuous} to require more mode switches than \emph{Threshold}, as users have no clear guidance about when to switch modes. This may cause them to oversteer or accept new suggestions inefficiently.
 \item[\textbf{H3:}] \emph{Continuous} and \emph{Threshold} cause lower perceived workload compared to \emph{Classic}. However, we expect \emph{Continuous} to cause a higher workload compared to \emph{Threshold}, as it requires constant evaluation of the visual feedback while \emph{Threshold} allows the user to relax until further notification.
\end{enumerate}

\subsection{Apparatus}
\label{sec:apparatus}
\noindent Developing and testing new concepts for a robotic arm involves inherent challenges associated with a real robot's physical bulk and complexity. Quickly changing the experimental setup, adding feedback components, or providing information to the user further complicate testing regimes. We created a 3D testbed environment for \ac{HRI} studies in \ac{VR} to address these challenges. This testbed contains a simulated robotic arm (a virtual model of the \emph{Kinova Jaco 2}) with multiple control mechanisms and a standardized pick-and-place task. Visual feedback mimics \ac{AR}, with directional cues registered in 3D space. A \emph{Meta Quest} motion controller is used as an input device to control the robotic arm. 

\begin{figure}[htbp]
    \centering
    \includegraphics[width=\linewidth]{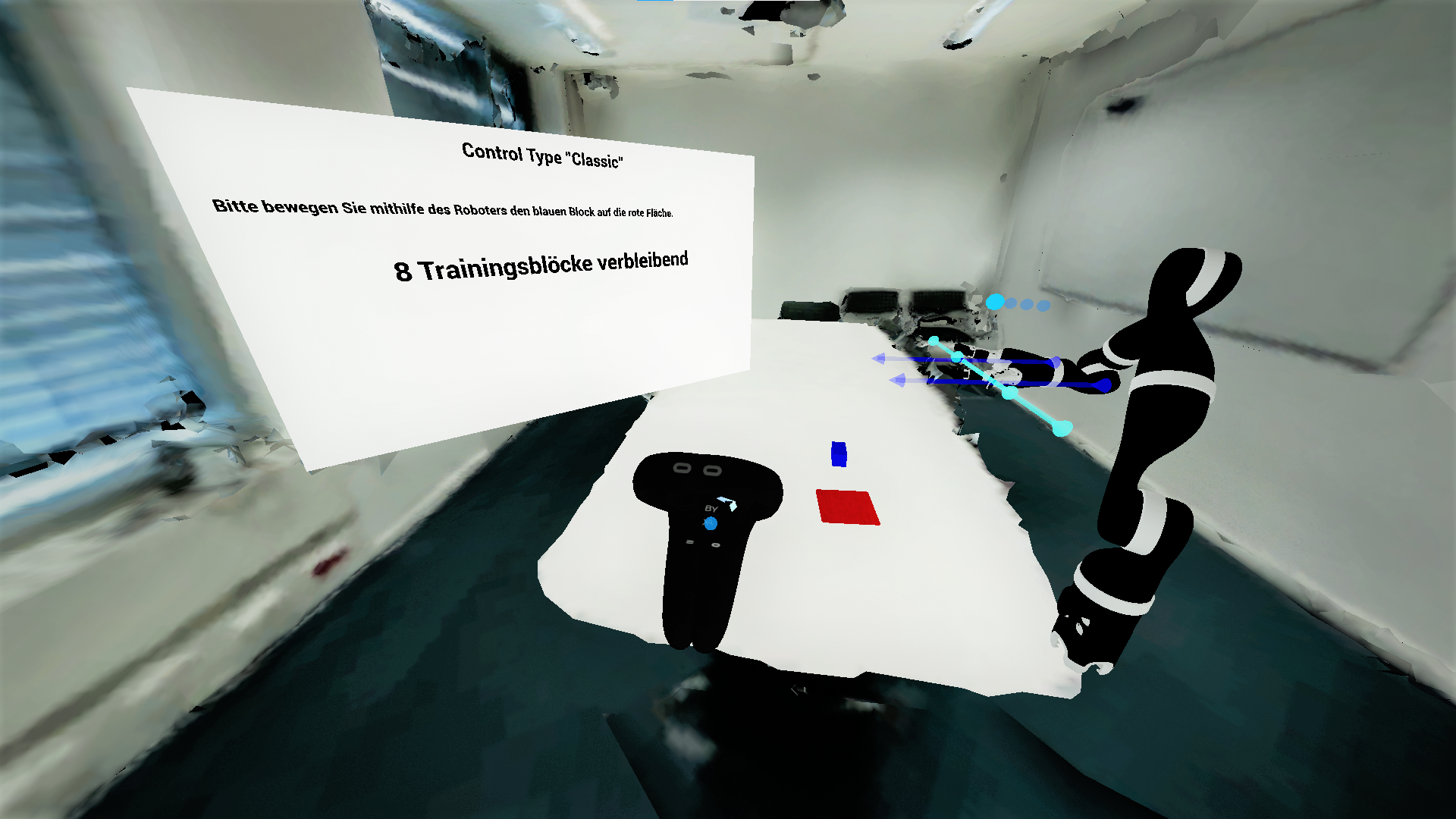}
    \caption{Virtual environment consisting of (left to right): a virtual canvas, the motion controllers, a table with the blue object and red target, and a \emph{Kinova JACO} with an arrow-based visualization}
    \label{fig:apparatus}
\end{figure}

Photogrammetry scans of an actual room were used to design the \ac{VR} environment, which was created using the \emph{Unreal Engine 4.27} and optimized for usage with a \emph{Meta Quest} \ac{VR} \ac{HMD} (see Figure~\ref{fig:apparatus}). During the study, user behavior was recorded with appropriate software on a \emph{Schenker XMG Key 17} laptop with \emph{Windows 10 64-bit} and \emph{Oculus Link} connected to the \ac{VR} headset.

For our implementation of the baseline control method \emph{Classic}, users cycled through four distinct modes to access all seven robot \acp{DoF}, as they are mapped on a two-\ac{DoF} joystick, such as the control-stick on a \emph{Meta Quest} motion controller: 

\begin{enumerate}
    \item X-Translation + Y-Translation
    \item Z-Translation + Roll
    \item Yaw + Pitch
    \item Open/Close fingers
\end{enumerate}

\noindent We illustrate the current mapping between the robot's \acp{DoF} and the input device through two arrows attached to the gripper. Light blue arrows indicate the robot's \ac{DoF} assigned to the first, dark blue arrows to the second input axis. Looking at the joystick in \ac{VR}, the same color-coded visualization is applied. 

Users press a button on the input device -- the A-Button of the \emph{Meta Quest} motion controller -- to switch between modes, cycling back to the first one at the end. 
Four blue spheres -- in contrast to the slanted cubes used in our \ac{ADMC} methods -- above the robotic arm's gripper indicate the total number of available and the currently active mode when users are not moving the robot. The sphere representing the active mode is bigger and brighter than the spheres of inactive modes. 

\subsection{Participants}
\label{sec:participants}
\noindent A total of 24 participants took part in our study (7 female, 17 male). The participants were aged 19 to 37, with a mean age of 26 years (SD~$=$~4.85 years). No one declared any motor impairments that might influence reaction times. Five participants had prior experience with controlling a robotic arm. Participants were recruited from a university campus and an online appointment form.

\subsection{Procedure}
\label{sec:procedure}
\noindent Utilizing the benefits of a standardized and portable \ac{VR} simulation environment, the study was conducted in multiple comparable physical localities. Before commencing, participants were fully informed about the project objective and the various tasks they had to complete. Every participant gave their full and informed consent to partake in the study, have video and audio recordings taken, and have all the relevant data documented. 

\noindent A study administrator observed the experiment on a laptop and briefed participants on using the hardware as well as the general functionalities of the study environment. Once set up, users followed command prompts embedded in the virtual simulation environment. For each of the three conditions, the following steps were performed: 

\begin{enumerate}
    \item Participants were given a written and standardized explanation of the control method used in the current condition.
    \item Participants conducted eight training trials for familiarization with the respective control method.
    \item Participants then conducted 24 measured trials.
    \item Interview and questionnaires.
\end{enumerate}

\noindent After completing all conditions, participants ranked the three control methods from most to least preferred and explained the reasoning behind their decision. The study concluded with a de-briefing. 
The average session lasted for 90 minutes and participants were compensated with 30 EUR.

\subsection{Experimental Task}
\label{sec:task}
The experimental task is based on our previous work and resembles a common pick-and-place scenario~\cite{Kronhardt.2022adaptOrPerish}. A blue object appears on a table in front of the participant, which signals the start of a trial. The user has to control the robot from its starting position to pick the object and place it on a red target surface, also located on the table. To change the \ac{DoF} mapping -- for trial fulfillment -- users could switch modes. Upon completion, the blue object disappears, and the robot automatically returns to the original starting position. A new blue object appears when this position is reached, and a new trial commences. For each trial, the position of the blue object is placed in one of eight possible locations spaced evenly around the red target surface. Each position occurred once during training and thrice during measured trials. However, the order of appearance was randomized. We used a neutral block shape rather than specific objects to avoid bias and ensure trial comparability.
\section{Results}
\noindent The study comprises 1,728 (24 participants $\times$ 3 control methods $\times$ 24 trials) measured trials. Training trials were excluded from the analysis.

We explored the distribution of the data through QQ-plots and either applied parametric \ac{RMANOVA} or non-parametric Friedman tests. For the latter, post-hoc pairwise comparisons using Wilcoxon signed-rank test with Bonferroni correction followed the omnibus test. Relevant effect sizes were calculated with r: $>$0.1 small, $>$0.3 medium, and $>$0.5 large effect.

\subsection{Task Completion Time}
\begin{figure}[htbp]
    \centering
    \includegraphics[width=\linewidth]{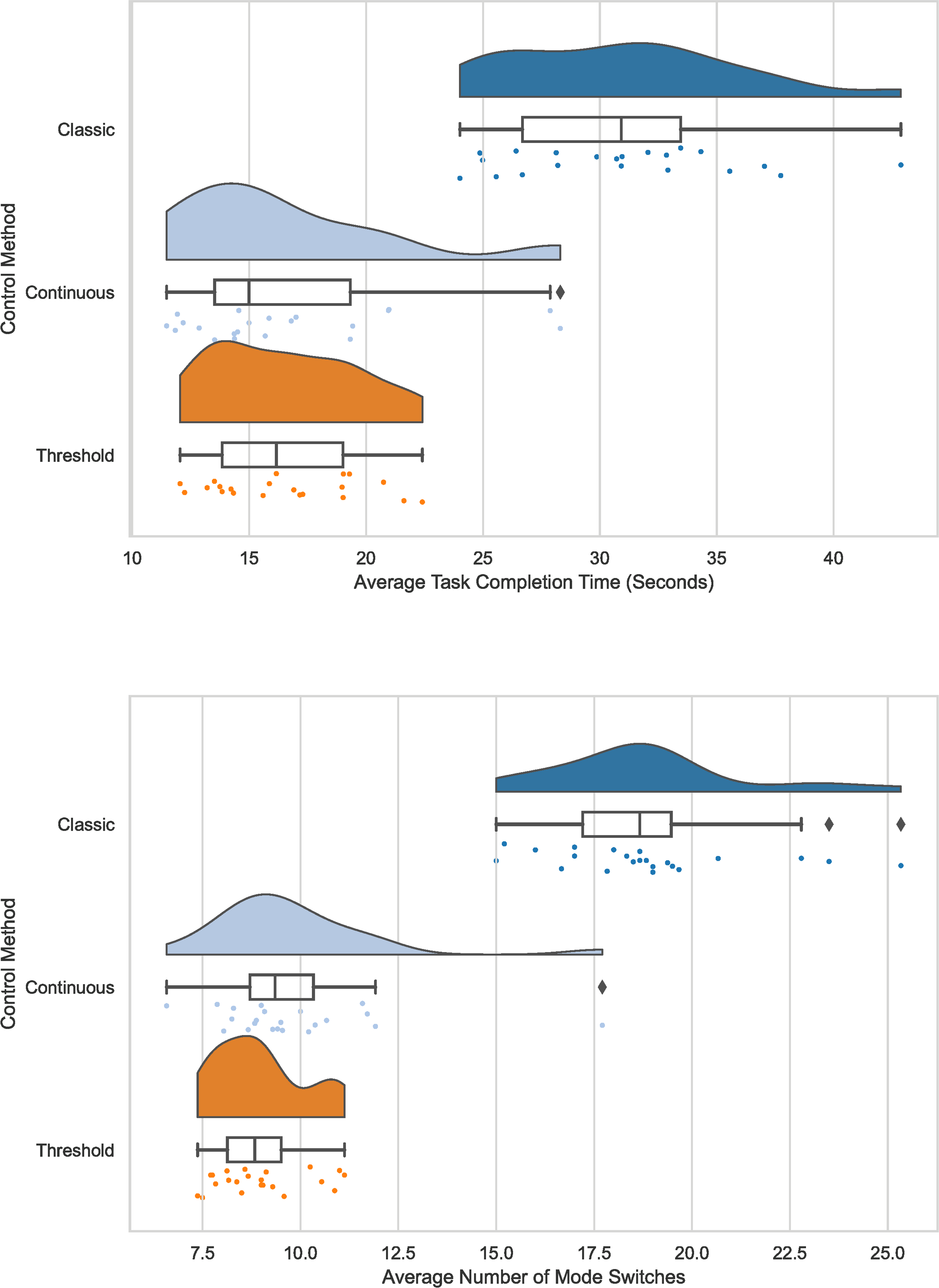}
    \captionsetup{justification=justified}
    \caption{Raincloud Plots for Average Task Completion Time and Mode Switches}
    \label{fig:raincloud}
\end{figure}

\noindent Mean task completion time calculated per participant and control method (see Fig.~\ref{fig:raincloud}) 
resulted in \emph{Threshold}~$=$~16.54s (SD~$=$~4.09s); \emph{Continuous}~$=$~16.61s (SD~$=$~4.77s); and \emph{Classic}~$=$~30.96s (SD~$=$~4.89s). Outliers [N~$=$~3] with average times $\geq$~2.2~$*$~IQR of the mean task completion time in at least one control method were excluded~\cite{Hoaglin1987}. The QQ-plot of the remaining 21 participants followed a normal distribution.

A \ac{RMANOVA} found a significant main effect (F(2, 36) = 130.92, p~$\leq$0.001). A post-hoc pairwise comparison (Bonferroni corrected) showed a significant difference between \emph{Continuous} and \emph{Classic} (p~$\leq$0.001) as well as between \emph{Threshold} and \emph{Classic} (p~$\leq$0.001). No significant difference was found between \emph{Continuous} and \emph{Threshold} (p~$\geq$0.999).

\subsection{Mode Switches}
\noindent We used a non-parametric Friedman test, as our data was not normally distributed, to determine differences between the average number of necessary mode switches between control methods. Two outliers -- based on $\geq$ 2.2 $*$ IQR of the mean value -- were excluded prior to further analysis. This resulted in mean numbers of mode switches for \emph{Threshold}~$=$~9.28 (SD~$=$~1.26); \emph{Continuous}~$=$~9.93 (SD~$=$~1.47); and \emph{Classic}~$=$~19.55 (SD~$=$~2.93) for N~$=$~22.
We found a significant main effect ($\chi^2$(2)~$=$~33.82, p~$\leq$0.001, N~$=$~22). Post-hoc pairwise comparisons showed a significant difference between \emph{Continuous} and \emph{Classic} (Z~$=$~$-$4.11, p~$\leq$0.001, r~$=$~0.62) as well as \emph{Threshold} and \emph{Classic} (Z~$=$~$-$4.11, p~$\leq$0.001, r~$=$~0.62). Again, we found no significant difference between the two \ac{ADMC} methods (Z~$=$~$-$1.51, p~$=$~0.131, r~$=$~0.28) (see Fig.~\ref{fig:raincloud}).


\subsection{Perceived Workload}
\noindent \ac{RTLX}~\cite{Hart.2006} scores [scale from 1 to 100] for all participants resulted in mean task load values of \emph{Threshold}~$=$~22.67 (SD~$=$~13.86); \emph{Continuous}~$=$~23.23 (SD~$=$~13.26); and \emph{Classic}~$=$~34.24 (SD~$=$~14.65). We applied a Friedman test which revealed a significant main effect for perceived task load: ($\chi^2$(2)~$=$~9.87, p~$=$~0.007, N~$=$~24). Post-hoc pairwise comparisons show significant differences between \emph{Continuous} and \emph{Classic} (Z~$=$~$-$3.03, p~$=$~0.002, r~$=$~0.44), \emph{Threshold} and \emph{Classic} (Z~$=$~$-$2.76, p~$=$~0.006, r~$=$~0.40), but not between \emph{Continuous} and \emph{Threshold} (Z~$=$~$-$0.21, p~$=$~0.830, r~$=$~0.03).

\subsection{Evaluation of Physical Assistive Devices}
\noindent The \ac{QUEAD} encompasses five individual scales (3 to 9 items each, 7-point Likert). Friedman tests for individual dimensions revealed significant main effects for \emph{Perceived Usefulness (PU)}, \emph{Perceived Ease of Use (PEU)}, \emph{Emotions (E)}, and \emph{Comfort (C)}, but not for \emph{Attitude (A)}. Post-hoc pairwise comparisons indicate significant differences between \emph{Continuous} and \emph{Classic} for \emph{PU}, \emph{PEU}, and \emph{C} as well as between \emph{Threshold} and \emph{Classic} for \emph{PU} and \emph{PEU} (refer to Table~\ref{tab:quead} for detailed scores).

\begin{table}[htbp]
    \centering
    \captionsetup{justification=justified}
    \caption{Statistics for individual \ac{QUEAD} dimensions: Perceived Usefulness (PU), Perceived Ease of Use (PEU), Emotions (E), Attitude (A), and Comfort (C). \label{tab:quead}}
    \footnotesize
    \begin{tabular}{p{2.7cm}ccccc}
        \toprule
        \textbf{}	& \textbf{PU} & \textbf{PEU} & \textbf{E} & \textbf{A} & \textbf{C} \\
        \midrule
        \multicolumn{6}{c}{\textbf{Descriptive Statistics}}                 \\
        \midrule
        $M_{\,Classic}$              & 4.98  & 4.87  & 5.00  & 4.81  & 5.65  \\
        $SD_{\,Classic}$             & 1.39  & 1.20  & 1.71  & 1.75  & 1.71  \\
        \midrule
        $M_{\,Continuous}$           & 5.68  & 5.80  & 5.90  & 5.42  & 6.44  \\
        $SD_{\,Continuous}$          & 1.05  & 1.04  & 1.25  & 1.48  & 0.78  \\
        \midrule
        $M_{\,Threshold}$            & 5.77  & 5.90  & 5.68  & 5.44  & 6.13  \\
        $SD_{\,Threshold}$           & 1.02  & 0.97  & 1.43  & 1.58  & 1.14  \\
        \midrule
        \multicolumn{6}{c}{\textbf{Friedman Tests}}                         \\
        \midrule
        $\chi^2$(2)                 & 7.49      & 15.22             & 7.20      & 1.76      & 6.39      \\
        $p$                           & 0.022   & $\leq$0.001   & 0.026   & 0.422     & 0.040   \\
        $N$                           & 24        & 24                & 24        & 24        & 24        \\
        \midrule
        \multicolumn{6}{c}{\textbf{Pairwise Comparisons}}                                               \\
        \midrule
        \multicolumn{3}{l}{Classic vs. Continuous}                            &           &           &           \\
        \midrule
        $|Z|$                         & 2.32      & 2.47              & 1.85      & ---       & 2.29      \\
        $p$                           & 0.021   & 0.014           & 0.064     & ---       & 0.022   \\
        $r$                           & 0.33      & 0.36              & 0.27      & ---       & 0.33      \\
        \midrule
        \multicolumn{3}{l}{Classic vs. Threshold}                            &           &           &           \\
         \midrule
        $|Z|$                         & 2.68      & 2.90              & 1.28      & ---       & 1.23      \\
        $p$                           & 0.007  & 0.003          & 0.202     & ---       & 0.220     \\
        $r$                           & 0.39      & 0.43              & 0.18      & ---       & 0.18      \\
        \midrule
        \multicolumn{3}{l}{Continuous vs. Threshold}                           &           &           &           \\
         \midrule
        $|Z|$                         & 0.62      & 0.38              & 1.03      & ---       & 1.70      \\
        $p$                           & 0.538     & 0.706             & 0.302     & ---       & 0.089     \\
        $r$                           & 0.09      & 0.05              & 0.15      & ---       & 0.25      \\
        \bottomrule
    \end{tabular}
\end{table}

\subsection{Individual Ranking}
\noindent Participants ranked the control methods in order of preference from 1 = \emph{favorite} to 3 = \emph{least favorite}. Mean values in ascending order are \emph{Continuous}~$=$~1.67; \emph{Threshold}~$=$~2.04; and \emph{Classic}~$=$~2.29.
A Friedman test revealed no significant main effect ({$\chi^2$(2)}~$=$~4.75, p~$=$~0.100, N~$=$~24).

\subsection{Qualitative Insights}
\label{sec:qualitativeInsights}
\noindent Overall, the open coding process led to the identification of five main themes, as discussed below.

\subsubsection{Familiarization}
While all three control methods included a training phase, comments suggest that in particular the \ac{ADMC} methods required familiarization. Here, participants felt the controls were sometimes \enquote{inverted} (P3) and wanted to \enquote{move the stick in the direction the arrow was pointing at} (P6). They also reported that \enquote{it takes a while to get used to} (P24), but \enquote{routine set in fast} (P18).

\subsubsection{Handling Adaptive \ac{DoF} Mapping Suggestions}
\label{sec:qihandling}
The study cohort showed a relatively uniform response to the two \ac{ADMC} methods with clear distinctions between \emph{Threshold} and \emph{Continuous}. In \emph{Threshold}, many participants \enquote{trusted the system} (P23) and switched to the new suggestion as soon as they perceived the multimodal indicator. They \enquote{did not have to think a lot} (P4) and \enquote{relied on what the suggestion says} (P7). This dependence on the system caused some to \enquote{draw a blank when something went wrong because [they] forgot they had other options} (P8). One participant even tried using the \emph{Threshold} control method with eyes closed, which \enquote{worked surprisingly well} (P7). 

In contrast, participants evaluated the suggestions in \emph{Continuous} more thoroughly, as they had to decide when to switch without the help of threshold-based indicators. Some participants waited for suggestions with relatively simple direction cues, such as \enquote{straight arrows} (P6, P16) as an indication to switch modes, while others trusted their \enquote{gut feeling} (P23). Uncertainties of \enquote{How do I approach this?} (P23) were more frequent in this control method than \emph{Threshold}. 
Participants dealt with problems in both \ac{ADMC} conditions in one of two ways to find alternative suggestions that better align with their needs. They cycled through the further offered suggestions for an alternative option or reversed their current movement direction until a different suggestion was offered.

\subsubsection{Visualization}
Overall, participants understood the different visualizations. Yet, difficulties arose in all three conditions relating to depth perception and understanding if the gripper is positioned correctly to pick or place the object. Some participants suggested a \enquote{laser pointer} (P16) to indicate the gripper's position above the table for improved depth perception. This is a known problem for robot teleoperation. In the past, researchers have suggested and explored \ac{AR} \emph{Visual Cues} to counter that, which include similar approaches as the ones mentioned by our participants~\cite{Arevalo-Arboleda2021,Arevalo.2021b.Interact}. 

Interestingly, some participants \enquote{manipulated} the second mode of \emph{Classic} (X- and Y-Translation) to mimic this effect, as that mode shows straight up- and downward pointing arrows as directional cues along the y-axis. 

\subsubsection{Multimodal Feedback}
As described above, most participants used \emph{Threshold} as intended, switching to the next suggestion when they received the multimodal feedback. However, some participants experienced the haptic and audio indicators as \enquote{irritating} (P20) or \enquote{weird and horrible} (P17). The poignant statement \enquote{If I had to do this for five more minutes, it would be too annoying.} (P7) reveals some participants' strong reactions to this control method. As a possible mitigation, one participant suggested implementing multiple thresholds of varying intensity instead of a singular one that \enquote{instantly beeps loudly at me and says 'Do this now!'} (P24).

\subsubsection{Control vs. Comfort}
Participants reported substantial differences in the level of control and comfort between \emph{Classic}, \emph{Continuous}, and \emph{Threshold}. By nature, \emph{Classic} offers the highest control level but requires participants to decide individually on every task step. In contrast, \emph{Threshold} allowed participants to perform tasks \enquote{entirely brainlessly} (P16) and only press \enquote{forward, then A, then forward, then A} (P17). Many participants expressed that they \enquote{felt too directed by [\emph{Threshold}]} (P8), attesting \emph{Continuous} a higher level of comfort or \enquote{freedom to experiment} (P24). Overall, participants described \emph{Continuous} as a reasonable compromise or \enquote{the golden middle} (P14) between the comfortable execution in \emph{Threshold} and the high level of control in \emph{Classic}.
\section{Discussion}
\label{sec:discussion}
\noindent Adaptive \ac{DoF} mapping controls have already been indicated to have benefits over classic methods~\cite{Goldau.2021petra,Kronhardt.2022adaptOrPerish}. 
Yet, research is still limited, and analysis of \emph{time-based dimensions} of directional cues is lacking. In this paper, we examined to what extent the two \ac{ADMC} methods, \emph{Continuous} and \emph{Threshold}, differ from the \emph{Classic} baseline -- and each other -- in terms of task completion time, necessary mode switches, perceived workload, and subjective assessment.

Significant results for all four metrics partially support our initial hypotheses. Most strikingly, \ac{ADMC} methods reduced task completion time (\emph{H1}) and mode switches (\emph{H2}) by 50\% respectively compared to \emph{Classic}. 
As previously suggested by Kim et al., this establishes that \ac{ADMC} methods lead to faster and less involved execution of pick-and-place tasks~\cite{Kim.2012}. These findings are in line with previous work~\cite{Goldau.2021petra}, underlining the benefits of \acp{ADMC} compared to \emph{Classic} controls.

In contrast to previous results~\cite{Kronhardt.2022adaptOrPerish}, our novel \ac{ADMC} methods were able to significantly lower task completion time and perceived workload compared to the \emph{Classic} method. The latter finding also partially supports \emph{H3}. This highlights that \acp{ADMC} which communicate the suggested recommendation to the user -- irrespective of timing -- were able to increase usability. 
Notably, the decreased workload of \acp{ADMC} is particularly meaningful as the end goal should be the smooth integration of robotic devices into people's lives and workflows, not to add stress. 

Turning to the second part of our analysis -- contrasting different time-based communication of feed-forward recommendations -- we found no significant differences in the four metrics between \emph{Continuous} and \emph{Threshold}. 
The lack of measurable differences between \emph{Continuous} and \emph{Threshold} implies that both discrete and continuous communication of movement suggestions allows users to use \ac{ADMC} methods efficiently. Insights gained by the results of the \ac{QUEAD} and our qualitative interviews corroborate these findings, while the latter also helped to provide a more distinguished analysis.

Overall, participants expressed a positive stance regarding the \ac{ADMC} methods. However, individual preferences vary greatly between \emph{Continuous} and \emph{Threshold}. While some participants preferred the higher level of control \emph{Continuous} allowed, others favored the comfortable execution possible with \emph{Threshold}. Consequently, future development of \ac{ADMC} methods should -- in accordance with Burkolter et al. -- include individualization options to increase comfort and end-user acceptance~\cite{Burkolter.2014customization}. Customizations would be particularly beneficial for \emph{Threshold}-based controls as participants repeatedly criticized the multimodal feedback. Allowing users to adjust the modalities, the signal intensity, and even the threshold itself may improve usability while still offering the advantages of \ac{ADMC}.

In contrast to expectations derived from our initial hypotheses, qualitative insights revealed that the \emph{Classic} control method could still be a valuable addition in specific situations. Participants felt an apparent lack of control when the \ac{ADMC} suggestions did not match their expectations. To improve usability, \ac{ADMC} methods could incorporate static suggestions for certain situations. A potential way to address this could be combining \ac{ADMC} and static suggestions using only the most common input-\acp{DoF}. 

However, further experimental studies are needed to disentangle exactly which factors shape personal preferences and how customizations or crossover methods can deliver the best results.

\subsection{Limitations}
\label{sec:limitations}
\noindent We explored the proposed \ac{ADMC} methods in a \ac{VR} simulation environment. While the usage of virtual simulations in industrial settings has been successfully established~\cite{Matsas.2015vrIndustry,Muller.2017vrIndustry,Tidoni.2017vrIndustry}, future work should confirm if our promising findings can be replicated in the real world with a physical robot.
\section{Conclusions}
\label{sec:conclusion}
\noindent Our \ac{ADMC} methods \emph{Continuous} and \emph{Threshold} are promising approaches to communicate proposed directional cues effectively. We extend our previous work~\cite{Kronhardt.2022adaptOrPerish} by demonstrating that \acp{ADMC} significantly reduce task completion time (1), the average number of necessary mode switches (2), and the perceived workload of the user (3). Further, we establish that \emph{Continuous} and \emph{Threshold} perform equally well in quantitative measures while qualitative insights reveal individual preferences. 

The observations of this study provide valuable implications for any \ac{HRI} researcher involved in designing novel \ac{ADMC} methods for human-robot collaborative settings. 
Future work should focus on disentangling quantitative and qualitative feedback of focus groups to develop optimal robot motion control methods, thus increasing usability, safety and -- ultimately -- end-user acceptance.

\section*{ACKNOWLEDGMENT}
\noindent This research is supported by the \textit{German Federal Ministry of Education and Research} (BMBF, FKZ: 16SV8563, 16SV8565). Our study is approved by the Ethics Committee of the \textit{Faculty of Business Administration and Economics of the University of Duisburg-Essen} with the ID: 2202007.

\bibliographystyle{IEEEtran}
\bibliography{bibliography}

\end{document}